\documentclass[12pt]{article}

\begin{document}
\begin{center}
{\bf MAXWELL'S THEORY ON NON-COMMUTATIVE SPACES AND QUATERNIONS}\\
\vspace{5mm}
 S.I. Kruglov \\
\vspace{5mm}
\textit{International Educational Centre, 2727 Steeles Ave. W, \# 202, \\
Toronto, Ontario, Canada M3J 3G9}
\end{center}

\begin{abstract}
The Maxwell theory on non-commutative spaces has been considered.
The non-linear equations of electromagnetic fields on
non-commutative spaces were obtained in the compact spin-tensor
(quaternion) form. It was shown that the plane electromagnetic
wave is the solution of the system of non-linear wave equations of
the second order for the electric and magnetic induction fields.
We have found the canonical and symmetrical energy-momentum
tensors and their non-zero traces. So, the trace anomaly of the
energy-momentum tensor was obtained in electrodynamics on
non-commutative spaces. It was noted that the dual transformations
of electromagnetic fields on non-commutative spaces are broken.
\end{abstract}

\section {Introduction}

The field models on non-commutative (NC) spaces are of great
interest now due to the recent development of the superstring
theory. It was shown that NC coordinates emerge naturally in the
perturbative version of the $D$-brane theory (low energy
excitations of a $D$-brane) with the presence of the external
background magnetic field [1]. So, non-commutative Yang-Mills
(NCYM) theories appear in the string theory and therefore they are
being widely investigated. The NC field theories have the same
degrees of freedom as effective commutative theories and,
therefore, there exists a map (the Seiberg-Witten map between NC
field theory and the corresponding commutative field theory)
between them. The simplest theory with the gauge group $U(1)$ is
QED and its prototype - NC quantum electrodynamics (NCQED). The
investigation of gauge theories on NC geometry leads to the
non-local interactions of fields due to the presence of higher
derivatives in the Lagrangian. The distinctive features of the NC
theory are the appearance of the dipole moments of particles at
one loop level and the violation of the CP-symmetry [2]. So, in
NCQED the ``electron" possesses the magnetic dipole moment which
contains the spin-independent term (proportional to the
non-commutative parameter $\theta$) and the electric dipole moment
violating the CP-symmetry, but it should be noted that the
CPT-symmetry remains unbroken [3]. At the charge conjugation the
theory transfers to the sector with $\theta\rightarrow -\theta$.
At one-loop level NCQED is a renormalizable [4,5] and asymptotic
free (the $\beta$-function is negative and is not $\theta$
dependent) theory [6]. Besides, the parameter $\theta$ does not
acquire the quantum corrections. The non-commutative version of a
standard model was considered in [2,7]. It should be mentioned
that NC field theories possess unitarity at the space-like
non-commutative tensor $\theta_{\mu\nu}$ ($\theta_{0j}=0$) [8,9].
There is infrared-ultraviolet (IR/UV) mixing in NC theories, and
if we remove the UV divergences at cut-off$\rightarrow\infty$, the
new IR divergences appear.

The assumption that coordinates do not commute was made a long
time ago [10] (see also [11]). The NC coordinates of the
corresponding spaces obey the following commutation relation

\begin{equation}
 \left[\widehat{x}_\mu,\widehat{x}_\nu \right]=i\theta_{\mu\nu} ,
\label{1}
\end{equation}

where the non-commutative parameter $\theta_{\mu\nu}$ possesses
the dimension of $(\mbox{length})^2$. It is implied that we have
ordinary commutative relations between coordinates
$\widehat{x}_\mu$ and the momentum $\widehat{p}_\mu$:
$\left[\widehat{x}_\mu,\widehat{p}_\nu
\right]=i\hbar\delta_{\mu\nu}$,
$\left[\widehat{p}_\mu,\widehat{p}_\nu \right]=0$. The
astro-physical bounds on the NC scale $\Lambda_{NC}$ [7,12] are
given by
\begin{equation}
 \theta_{\mu\nu}=\frac{1}{\Lambda_{NC}^2}\epsilon_{\mu\nu},
 \hspace{0.3in}\Lambda_{NC}\geq10^3~\mbox{GeV} ,
\label{2}
\end{equation}
where $\epsilon_{\mu\nu}$ is a dimensionless antisymmetric tensor,
$\epsilon_{\mu\nu}=-\epsilon_{\nu\mu}$. The parameter
$\theta_{\mu\nu}$ is extremely small, and, therefore, observable
effects can emerge only at the cosmological scale (of the order of
the Plank length), i.e. at high energy. As $\theta_{\mu\nu}$ is a
constant tensor, the Lorentz symmetry is broken for field theories
on NC geometry. It was noted in [7] that at the replacement
\begin{equation}
 x_i=\widehat{x}_i +\frac{1}{2\hbar}\theta_{ij}\widehat{p}_j,
 \hspace{0.3in}p_j=\widehat{p}_j ,
\label{3}
\end{equation}
we arrive at the standard commutation relationships:
$\left[x_i,x_j \right]=0$, $\left[x_i,p_j
\right]=i\hbar\delta_{ij}$, $\left[p_i,p_j \right]=0$. The
non-local character of field theories on NC geometry follows from
Eq. (3).

The field operators $\widehat{A}(\widehat{x})$ in field theories
on NC geometry are the functions of $\widehat{x}_\mu$. After the
Fourier transformation we have
\begin{equation}
 \widehat{A}(\widehat{x})=\frac{1}{(2\pi)^4}\int d^4 p\exp(ip_\mu
 \widehat{x}_\mu )A(p),\hspace{0.3in}A(p)=\int d^4 x\exp(-ip_\mu
 x_\mu)A(x) .
\label{4}
\end{equation}
Then the product of two field operators
$\widehat{A}(\widehat{x})$, $\widehat{B}(\widehat{x})$ can be
represented as
\[
 \widehat{A}(\widehat{x})\widehat{B}(\widehat{x})=
 \int\frac{d^4 p}{(2\pi)^4}\int\frac{d^4 k}{(2\pi)^4}
 \exp\left\{i(p_\mu +k_\mu )x_\mu -\frac 12 p_\mu k_\nu
 \left[\widehat{x}_\mu,\widehat{x}_\nu \right]\right\}A(p)B(k)
\]
\vspace{-8mm}
\begin{equation}
\label{5}
\end{equation}
\vspace{-8mm}
\[
=\left[\exp\left(\frac i2
\theta_{\mu\nu}\partial_\mu\partial'_\nu\right)
A(x)B(x')\right]\mid_{x=x'} ,
 \]
 where $\partial_\mu=\partial/\partial x_\mu$,
 $\partial'_\nu=\partial/\partial x'_\nu$.
 Thus we come to the Weil-Moyal correspondence [13,14]:
 \begin{equation}
\widehat{A}(\widehat{x})\widehat{B}(\widehat{x})\longleftrightarrow
A(x)\star B(x) ,
 \label{6}
\end{equation}
where the star-product ($\star$-product) is given by
\begin{equation}
A(x)\star B(x)=\left[\exp\left(\frac i2
\theta_{\mu\nu}\partial_\mu\partial '_\nu\right)
A(x)B(x')\right]\mid_{x=x'} .
 \label{7}
\end{equation}
Using Eq. (7) it is easy to check that quadratic terms in the
actions of NC theories (kinetic terms) coincide with that of their
commutative versions, i.e., propagators are identical. The
star-product also satisfies the associative low: $(F\star G)\star
H=F\star (G\star H)$.

We use the system of units $\hbar=c=1$, $e^2/4\pi=1/137$, $e>0$.

\section{Field equations}

The free Maxwell action on NC space [5] is given by
\begin{equation}
 S=-\frac 14 \int d^4 x F_{\mu \nu }\star F_{\mu \nu }=
-\frac 14 \int d^4 x \widehat{F}^2_{\mu \nu } ,
 \label{8}
\end{equation}
were non-commutative strength $\widehat{F}_{\mu \nu }$ reads
\begin{equation}
 \widehat{F}_{\mu \nu }=\partial _\mu A_\nu -\partial _\nu A_\mu
 -ie\left [A_\mu,A_\nu\right ]_M ,
\label{9}
\end{equation}
and the Moyal bracket is
\begin{equation}
 \left [A_\mu,A_\nu\right ]_M=A_\mu \star A_\nu -A_\nu \star
 A_\mu .
\label{10}
\end{equation}
 The Seiberg-Witten expansion to the first order in $\theta_{\mu\nu}$
 [5] gives
\[
\widehat{A}_\mu=A_\mu-\frac 12 \theta_{\alpha\beta}A_\alpha
\left(\partial_\beta A_\mu +F_{\beta\mu}\right) ,
\]
\vspace{-8mm}
\begin{equation}
\label{11}
\end{equation}
\vspace{-8mm}
\[
 \widehat{F}_{\mu \nu }=F_{\mu \nu }+\theta_{\alpha\beta}
 F_{\mu \alpha}F_{\nu \beta}-\theta_{\alpha\beta}A_\alpha\partial_\beta
 F_{\mu \nu } ,
\]
with $e$ absorbed in $\theta_{\alpha\beta}$. The field strength
tensor corresponding to the commutative Maxwell theory is given by

 \begin{equation}
 F_{\mu \nu }=\partial _\mu A_\nu -\partial _\nu A_\mu ,
\label{12}
\end{equation}

with the vector-potential of the electromagnetic field $A_\mu$,
the electric field $E_i =iF_{i4}$ and the magnetic induction field
$B_{i} =\epsilon_{ijk}F_{jk}$ ($\epsilon_{123}=1$). The Lagrangian
within four-dimensional divergences (see [5]) is
\begin{equation}
 {\cal L}=-\frac14F^2_{\mu \nu}+\frac18\theta_{\alpha\beta}
 F_{\alpha\beta}F^2_{\mu \nu}-\frac12\theta_{\alpha\beta}
 F_{\mu \alpha}F_{\nu \beta}F_{\mu \nu}+{\cal O}(\theta^2)+A_\mu J_\mu ,
\label{13}
\end{equation}
where we added the external four-current $J_\mu$ and took into
consideration that the term $A_\mu\star J_\mu$ in the Maxwell
Lagrangian on NC spaces coincides within four-divergences with
$A_\mu J_\mu$. The Lagrangian (13) can also be cast in the form of
\begin{equation}
 {\cal L}=\frac12 \left( {\bf E}^2-{\bf B}^2 \right)\left[1+({\bf
 \theta}\cdot{\bf B})\right]-\left({\bf \theta}\cdot{\bf E}\right)\left({\bf
 E}\cdot{\bf B}\right)
 +{\cal O}(\theta^2)+A_\mu J_\mu ,
\label{14}
\end{equation}
where $\theta_i=(1/2)\epsilon_{ijk}\theta_{jk}$, $\theta_{i4}=0$.
It is seen from Eq. (14) that terms containing the non-commutative
parameter $\theta$ violate CP - symmetry. Using the Lagrange-Euler
equations
 \begin{equation}
 \partial_{\mu}\frac{{\partial\cal L}}{\partial\left(\partial
 _\mu A_\nu\right)} -\frac{{\partial\cal L}}{\partial A_\nu}=0 ,
\label{15}
\end{equation}
 we obtain from Eq. (13) field equations (equations of motion)
\[
\partial_\mu F_{\nu\mu}+\frac12\theta_{\alpha\beta}\partial_\mu
\left( F_{\mu \nu}F_{\alpha\beta}\right)+\frac14\theta_{\mu \nu}
\partial_\mu\left(F^2_{\alpha\beta}\right)
\]
\vspace{-8mm}
\begin{equation}
\label{16}
\end{equation}
\vspace{-8mm}
\[
-\theta_{\nu\beta}\partial_\mu\left(F_{\alpha
\beta}F_{\mu\alpha}\right)+\theta_{\mu\beta}\partial_\mu\left(F_{\alpha
\beta}F_{\nu\alpha}\right)-\theta_{\alpha\beta}\partial_\mu\left(
F_{\mu\alpha}F_{\nu\beta}\right)=J_\nu .
\]
The non-linear equations (16) may be cast as follows [15]:
\begin{equation}
\frac{\partial}{\partial t}{\bf D}-\mbox{rot} {\bf H}=-{\bf
J},\hspace{0.3in} \mbox{div}{\bf D}=\rho , \label{17}
\end{equation}

where $(\mbox{rot} {\bf H})_{i} =\epsilon_{ijk}\partial_{j}H_{k}$
and $\mbox{div}{\bf D}=\partial_{i}D_{i}$; ${\bf J}$ is the vector
of a current and $\rho$ is a charge density, $J_\mu$=$({\bf
J},i\rho )$. The displacement (${\bf D}$) and magnetic (${\bf H}$)
fields are given by

\begin{equation}
 {\bf D}={\bf E}+{\bf d},\hspace{0.3in}{\bf d}=({\bf \theta}\cdot
 {\bf B}){\bf E}-({\bf \theta}\cdot {\bf E}){\bf B}-
 ({\bf E}\cdot {\bf B}){\bf \theta} ,
\label{18}
\end{equation}

\begin{equation}
 {\bf H}={\bf B}+{\bf h},\hspace{0.3in}{\bf h}=({\bf \theta}\cdot
 {\bf B}){\bf B}+({\bf \theta}\cdot {\bf E}){\bf E}-
 \frac{1}{2}\left({\bf E}^2- {\bf B}^2\right){\bf \theta} .
\label{19}
\end{equation}
Here the scalar products of vectors are introduced, $({\bf
\theta}\cdot {\bf E})=\theta_i E_i$, and so on. The other equation
following from Eq. (12) is
 \begin{equation}
 \partial_\mu \widetilde{F}_{\mu \nu }=0 ,
\label{20}
\end{equation}
where $\widetilde{F}_{\mu\nu}=(1/2)\varepsilon _{\mu \nu \alpha
\beta }F _{\alpha \beta}$ is the dual tensor, $\varepsilon _{\mu
\nu \alpha \beta }$ is an antisymmetric tensor Levy-Civita;
$\varepsilon _{1234}=-i$. Eq. (20) is rewritten as
\begin{equation}
\frac {\partial}{\partial t}{\bf B}+\mbox{rot}{\bf
E}=0,\hspace{0.3in} \mbox{div}{\bf B}=0 . \label{21}
\end{equation}

Let us obtain the second order equations for electric and magnetic
fields. Such wave equations are convenient for studying the
propagation of electromagnetic fields. Applying the operator rot
to the first equation of (17) with the help of the equality
$\mbox{rot rot}{\bf H}=\mbox{grad div}{\bf H}-\triangle{\bf H}$
[$(\mbox{grad})_i\equiv\partial/\partial x_i$,
$\triangle\equiv\partial^2/(\partial x_i)^2$], replacing in this
equation $\mbox{rot}{\bf E}$ from Eq. (21) and taking into account
Eqs. (18), (19), we find
\begin{equation}
\triangle{\bf B}-\frac{\partial^2}{(\partial t)^2}{\bf
B}+\triangle{\bf h}-\mbox{grad div}{\bf
h}+\frac{\partial}{\partial t}\mbox{rot}{\bf d}=-\mbox{rot}{\bf
J}. \label{22}
\end{equation}
We repeat this procedure, starting with the first equation of (21)
and taking into consideration Eqs. (17)-(19), one obtains
\begin{equation}
\triangle{\bf E}-\frac{\partial^2}{(\partial t)^2}{\bf
E}-\frac{\partial^2}{(\partial t)^2}{\bf d}+\mbox{grad div}{\bf
d}+\frac{\partial}{\partial t}\mbox{rot}{\bf
h}=\frac{\partial}{\partial t}{\bf J}+\mbox{grad}\rho. \label{23}
\end{equation}
It is easy to verify that electric and magnetic induction fields
in the form of plane electromagnetic waves, such as
\begin{equation}
{\bf E}= {\bf E}_0\exp(ik_\mu x_\mu),\hspace{0.3in} {\bf B}=
\left({\bf n}\times {\bf E}_0\right)\exp(ik_\mu x_\mu),\label{24}
\end{equation}
where ${\bf n}={\bf k}/k_0$, $k_\mu=({\bf k},ik_0)$ [$({\bf
n}\times{\bf E}_0)_i=\epsilon_{ijk}n_jE_{0k}$ is the vector
product], are the solutions of the linear wave equations of
classical electrodynamics as well non-linear Eqs. (22), (23) at
${\bf J}=0$, $\rho=0$.
 It was shown in [15] that in the theory under
consideration the velocity of propagation when transverse to a
background magnetic induction field differs from $c$. This is a
consequence of the non-linearity of field equations. But this
effect of the photon propagation is very small due to the
smallness of the non-commutative parameter $\theta$.

It is easy to see that free NC Maxwell's equations (16), (20) at
$J_\mu=0$ are not invariant under the dual transformations of
electromagnetic fields [16]
\[
F'_{\mu \nu }=F_{\mu \nu }\cos\alpha-\widetilde{F}_{\mu \nu
}\sin\alpha ,
\]
\vspace{-8mm}
\begin{equation}
\label{25}
\end{equation}
\vspace{-8mm}
\[
\widetilde{F}'_{\mu\nu}=\widetilde{F}_{\mu\nu}\cos\alpha+
F_{\mu\nu}\sin\alpha .
\]
The terms containing the parameter $\theta_{\mu\nu}$ in Eq. (16)
violate the dual symmetry (25), and the condition
$\theta_{\mu\nu}=0$ (at $J_\mu =0$) recovers the dual symmetry of
standard Maxwell's equations. It should be noted that dual
transformations (25) describe the symmetry of the polarization
space. For example, in the case of the plane electromagnetic wave,
Eqs. (24), these transformations rotate the polarization axes
around the wave vector.

\section {Energy and momentum of electromagnetic field on NC spaces}

Multiplying Eq. (17) by ${\bf E}$, Eq. (21) by ${\bf H}$, and
adding them, we find
\begin{equation}
{\bf E}\frac {\partial {\bf D}}{\partial t}+ {\bf H}\frac
{\partial {\bf B}}{\partial t}=-({\bf J}\cdot{\bf E})-\mbox{div}
({\bf E}\times{\bf H}) , \label{26}
\end{equation}

Using Eq. (18), (19) one can represent
\begin{equation}
{\bf E}\frac {\partial {\bf D}}{\partial t}+ {\bf H}\frac
{\partial {\bf B}}{\partial t}=\frac {\partial {\cal E}}{\partial
t} ,
 \label{27}
\end{equation}
where the energy density of the electromagnetic field is given by
\begin{equation}
{\cal E}=\frac {{\bf E}^2 +{\bf B}^2}{2}\left[1+({\bf \theta}\cdot
{\bf B})\right]-\left({\bf E}\cdot{\bf B}\right) \left({\bf
\theta}\cdot{\bf E}\right) . \label{28}
\end{equation}
With the help of Eqs. (26)-(28), the conservation law of the
energy-momentum reads [17]
\begin{equation}
\frac {\partial {\cal E}}{\partial t}=-({\bf J}\cdot{\bf
E})-\mbox{div} {\bf P},\hspace{0.3in} {\bf P}={\bf E}\times{\bf H}
, \label{29}
\end{equation}
where ${\bf P}$ is the vector of the momentum density of the
electromagnetic field, so that the four-vector of the
energy-momentum is $P_\mu=({\bf P},i{\cal E})$. It follows from
Eq. (29) that at ${\bf J}=0$ the continuity equation $\partial_\mu
P_\mu =0$ holds. The energy density (28) and the momentum density
of the electromagnetic field with the accuracy of
$\textit{O}(\theta^2)$ and  using Eqs. (18),(19) may be
represented as
\[
{\cal E}=\frac {{\bf D}^2 +{\bf H}^2}{2}-\left({\bf \theta}\cdot
{\bf B} \right){\bf B}^2 ,
\]
\vspace{-8mm}
\begin{equation}
\label{30}
\end{equation}
\vspace{-8mm}
\[
{\bf P}=\left[1+({\bf \theta}\cdot{\bf B})\right]({\bf
E}\times{\bf B})+\frac12 \left({\bf B}^2-{\bf E}^2\right)({\bf
E}\times {\bf \theta}) .
\]
Now we use the general expression for the canonical conservative
energy-momentum tensor [17]:
\begin{equation}
 T_{\mu\nu}^{can}=\left(\partial_\nu A_\alpha\right)
 \frac{\partial{\cal L}}{\partial\left(\partial
 _\mu A_\alpha\right)} -\delta_{\mu\nu}{\cal L} ,
\label{31}
\end{equation}
so that $\partial_\mu T_{\mu\nu}^{can}=0$. To get the
gauge-invariant energy-momentum tensor, we can explore the
transformation [17]:
\begin{equation}
 T_{\mu\nu}=T_{\mu\nu}^{can}+ \Lambda_{\mu\nu} ,
\label{32}
\end{equation}
where the function $\Lambda_{\mu\nu}$ obeys the equation
$\partial_\mu \Lambda_{\mu\nu}=0$ due to equations of motion.
Indeed, it follows from Eq. (32) that the tensor $T_{\mu\nu}$ is
conservative, i.e., $\partial_\mu T_{\mu\nu}=\partial_\mu
T_{\mu\nu}^{can}=0$. It should be noted that in classical linear
electrodynamics this transformation (32) is used to obtain the
symmetric energy-momentum tensor. We may choose the following
function
\begin{equation}
 \Lambda_{\mu\nu}=-\left(\partial_\alpha A_\nu\right)
 \frac{\partial{\cal L}}{\partial\left(\partial
 _\mu A_\alpha\right)} .
\label{33}
\end{equation}
It is easy to verify that the function $\Lambda_{\mu\nu}$, (33),
meets the requirement $\partial_\mu \Lambda_{\mu\nu}=0$ due to the
Lagrange-Euler equation (15) (at $J_\mu=0$ the equality
$\partial{\cal L}/\partial A_\nu=0$ holds) and the fact that
$\partial{\cal L}/\partial\left(\partial_\mu A_\alpha\right)$ is
anti-symmetric in indexes $\mu$ and $\alpha$. With the help of
Eqs. (13), (31)-(33) we obtain the gauge-invariant conservative
energy-momentum tensor of electromagnetic fields (at $J_\mu=0$) on
NC spaces
\[
T_{\mu\nu}=-F_{\mu\alpha}F_{\nu\alpha}\left(1-\frac12\theta_{\gamma\beta}
 F_{\gamma\beta}\right)+\frac14
 \theta_{\mu\alpha}F_{\nu\alpha}F^2_{\rho\beta}
\]
\vspace{-8mm}
\begin{equation}
\label{34}
\end{equation}
\vspace{-8mm}
\[
-\theta_{\mu\beta}F_{\gamma\nu} F_{\rho\beta}F_{\gamma\rho}-\left(
F_{\mu\alpha}F_{\nu\gamma}+F_{\nu\alpha}F_{\mu\gamma}\right)
\theta_{\alpha\beta}F_{\gamma\beta}-\delta_{\mu\nu}{\cal L} .
\]
The tensor (34) is still non-symmetric, however in the limit
$\theta\rightarrow 0$, for classical electrodynamics, the
energy-momentum tensor (34) becomes symmetric. From Eq. (34) we
obtain the components of the energy-momentum tensor
\[
T_{44}={\cal E} ,\hspace{0.3in}T_{m4}=-iP_m ,
\]
\[
T_{4m}=-i\epsilon_{mnk}\left\{E_n B_k\left[1+({\bf
\theta}\cdot{\bf B}) \right]+({\bf E}\cdot{\bf B})B_n
\theta_k\right\} ,
\]
\begin{equation}
T_{mn}=E_m E_n +B_m B_n -\frac12 \delta_{mn}\left( {\bf E}^2+{\bf
B}^2\right) +\left( {\bf \theta}\cdot{\bf B}\right)\left(2E_m E_n
+B_m B_n\right)
\label{35}
\end{equation}
\[
+\frac12\left( {\bf E}^2+{\bf B}^2\right)\left[B_m \theta_n -
2\delta_{mn}\left( {\bf \theta}\cdot{\bf B}\right)\right]-\left(
{\bf \theta}\cdot{\bf E}\right)E_n B_m
\]
\[
-\left( {\bf E}\cdot{\bf B}\right)\left[\theta_m E_n+\theta_n E_m
- \delta_{mn}\left( {\bf \theta}\cdot{\bf E}\right)\right]
-\left({\bf E}\times{\bf \theta}\right)_m\left({\bf B}\times{\bf
E}\right)_n .
\]
Eqs. (35) clearly show that the trace of  the energy-momentum
tensor does not equal zero, and is given by
\begin{equation}
T_{\mu\mu}=\left( {\bf \theta}\cdot{\bf B}\right)\left({\bf
E}^2-{\bf B}^2\right)-2\left( {\bf \theta}\cdot{\bf
E}\right)\left({\bf E}\cdot{\bf B}\right) .
 \label{36}
\end{equation}
Non-zero value of the energy-momentum tensor trace (36) indicates
the anomaly in the case of the NC electrodynamics at the classical
level. This reflects the fact that the classical conformal
invariance is broken. The violation of conformal invariance, or
the trace anomaly, relates to the violation of the Lorentz
invariance in NC space.

In order to find the symmetric tensor of the energy-momentum, we
will explore the general procedure of the curve coordinate system
usage [17]. In the curve space-time the Lagrangian (13) with the
accuracy of ${\cal O}(\theta^2)$, and at $J_\mu=0$, reads
\[
 {\cal L}=-\frac14F_{\mu \nu}F_{\alpha\beta}g^{\mu\alpha}
 g^{\nu\beta}\left(1+ \frac12\theta_{\gamma\delta}
 F_{\sigma\rho}g^{\gamma\sigma}g^{\delta\rho}\right)
\]
\vspace{-8mm}
\begin{equation}
\label{37}
\end{equation}
\vspace{-8mm}
\[
 - \frac12\theta_{\alpha\beta}F_{\mu \gamma}F_{\nu \delta}F_{\rho
 \sigma}g^{\alpha\gamma}g^{\beta\delta}g^{\mu\rho}g^{\nu\sigma} .
\]
We notice also that Eq. (37) looks like the covariant expression,
but the covariance is broken because the variable
$\theta_{\mu\nu}$ is not transformed as the second rank tensor at
the Lorentz transformations. Therefore the action corresponding to
the Lagrangian (37) is not a scalar at the transformations of the
metric $g'_{\mu\nu}=g_{\mu\nu}+\delta g_{\mu\nu}$. Consequently,
the conservation of the symmetrical energy-momentum tensor
obtained by variation of the Lagrangian (37) on the metric tensor
is questionable. Using the general formula for the symmetric
energy-momentum tensor (in the case when ${\cal L}$ does not
depend on $\partial_\alpha g^{\mu\nu})$ [17]
\[
T^{sym}_{\mu\nu}=\frac{2}{\sqrt{-g}}\frac{\partial\sqrt{-g}{\cal
L}}{\partial g^{\mu\nu}} ,
\]
 and varying the Lagrangian (37) on the metric tensor
$g^{\mu\nu}$, with the help of equation $g=-1$ for the Minkowski
space, we arrive at the symmetric energy-momentum tensor:
\begin{equation}
T^{sym}_{\mu\nu}=T_{\mu\nu}+\frac14 \theta_{\nu\alpha}
 F_{\mu\alpha}F^2_{\rho\beta}
-\theta_{\nu\beta}F_{\gamma\mu} F_{\rho\beta}F_{\gamma\rho} ,
\label{38}
\end{equation}
where the conservative tensor $T_{\mu\nu}$ is given by Eq. (34).
It is easy to check with the help of Eqs. (28), (29), (38) that
the equations $T^{sym}_{44}={\cal E}$, $T^{sym}_{m4}=-iP_m $ are
valid. From Eq. (38) we also find the spacial components of the
symmetric energy-momentum tensor (the stress tensor)
\[
T^{sym}_{mn}=E_m E_n +B_m B_n -\frac12 \delta_{mn}\left( {\bf
E}^2+{\bf B}^2\right) +\left( {\bf \theta}\cdot{\bf
B}\right)\left(3E_m E_n +B_m B_n\right)
\]
\[
+\frac12\left( {\bf E}^2+{\bf B}^2\right)\left[B_m
\theta_n+\theta_m B_n - 3\delta_{mn}\left( {\bf \theta}\cdot{\bf
B}\right)\right]
\]
\vspace{-8mm}
\begin{equation}
\label{39}
\end{equation}
\vspace{-8mm}
\[
-\left( {\bf E}\cdot{\bf B}\right)\left[\theta_m E_n+\theta_n E_m
- \delta_{mn}\left( {\bf \theta}\cdot{\bf E}\right)\right]-\left(
{\bf \theta}\cdot{\bf E}\right)\left(E_m B_n+E_n B_m\right)
 \]
\[
-\left({\bf E}\times{\bf \theta}\right)_m\left({\bf B}\times{\bf
E}\right)_n-\left({\bf E}\times{\bf \theta}\right)_n\left({\bf
B}\times{\bf E}\right)_m .
\]
From Eqs. (38), (39) we find the trace of the symmetric
energy-momentum tensor:
\begin{equation}
T^{sym}_{\mu\mu}=2\left( {\bf \theta}\cdot{\bf B}\right)\left({\bf
E}^2-{\bf B}^2\right)-4\left( {\bf \theta}\cdot{\bf
E}\right)\left({\bf E}\cdot{\bf B}\right) .
 \label{40}
\end{equation}
The trace of the symmetric energy-momentum tensor (40) is two
times greater than the trace of the conservative energy-momentum
tensor (36), i.e. $T^{sym}_{\mu\mu}=2T_{\mu\mu}$. It should be
noted that the trace anomaly contributes to the cosmological
constant. As a result the trace anomaly might be the source of
significant period of inflation in the early universe.

When two Lorentz invariants - $I_1\equiv{\bf E}^2-{\bf B}^2$ and
$I_2\equiv({\bf E}\cdot{\bf B})$ equal zero, i.e., $I_1=I_2=0$, in
the case of the electromagnetic waves, Eq. (24), the trace anomaly
vanishes, $T_{\mu\mu}=T^{sym}_{\mu\mu}=0$.  For classical
electrodynamics at $\theta=0$ we arrive at the known result that
the trace anomaly is absent for any electromagnetic fields,
$T_{\mu\mu}^{(\theta=0)}=0$.

\section{Spin-tensor form of equations}

Sometimes algebraic methods allow us to obtain results in the
simplest way. Multiplying Eqs. (21) by $i$, and adding Eqs. (17),
we can write
\begin{equation}
\partial_k f_k=\rho ,
\label{41}
\end{equation}
\begin{equation}
-i\epsilon _{kmn}\partial _m k_n-\frac{\partial f_k}{\partial
t}=J_k , \label{42}
\end{equation}

where $f_k=D_k+iB_k$, $k_n=E_n+i H_n$. Multiplying Eq. (42) by
$\tau _k$, and taking into account the properties of Pauli's
matrices, $\tau_k$ (see Appendix), one arrives at
\begin{equation}
-\tau _p\tau _k\partial _p k_k-i\partial _4\tau _lf_l=\tau _k J_k-
\partial _m k_m ,
\label{43}
\end{equation}

In accordance with Cartan's ideas [18], for every vector we can
construct a $2\times 2-$matrix $X$ (or $\overline{X}$) as follows
\[
X=x_\mu \tau _\mu ,\hspace{0.3in}\tau _\mu =\left( \tau _k,\tau
_4\right) ,
\]
\vspace{-8mm}
\begin{equation}
\label{44}
\end{equation}
\vspace{-8mm}
\[
\overline{X}=x_\mu \overline{\tau }_\mu
,\hspace{0.3in}\overline{\tau }_\mu =\left( -\tau _k,\tau
_4\right) ,
\]
where $\tau _4=i\tau _0$. With the help of Eqs. (17)-(21), Eq.
(43) can be cast in the form of
\begin{equation}
\nabla F+\frac12\left( \nabla G +G^{+} \overleftarrow{
\nabla}\right)=-J , \label{45}
\end{equation}
where $\nabla =\tau_\mu \partial_\mu$, $G=g_m \tau_m $,
$g_m=d_m+ih_m$, $J=J_\mu \tau_\mu$, $J_4=i\rho$, $F=f_m \tau_m$,
$f_4=0$; the matrix-differential operator $\overleftarrow{\nabla}$
acts on the left standing function, $G^{+}$ is Hermitian
conjugated matrix. We notice that the complex vector ${\bf
g}\equiv{\bf d}+i{\bf h}$ can be represented in the compact form:
\[
{\bf g}=i\left({\bf \theta}\cdot{\bf v}^{*}\right){\bf
v}-\frac{i}{2} {\bf \theta}\left({\bf v}^{*}\right)^2 ,
\]
where ${\bf v}={\bf E}+i{\bf B}$, ${\bf v}^{*}={\bf E}-i{\bf B}$.
  The spin-tensor $F=f_\mu \tau _\mu$ may be defined
through the potential as follows
\begin{equation}
F=-\overline{\nabla}A , \label{46}
\end{equation}
where $A=A_\mu \tau _\mu $, $\overline{\nabla}=\overline{\tau
}_\mu\partial _\mu $. From Eq. (46) we arrive at Eq. (12).

The spin-tensor form of NC Maxwell's equations (45) is equivalent
to the quaternion form as the quaternion algebra can be realized
through the Pauli matrices (see Appendix). At $\theta_{\mu\nu}=0$
we arrive from Eq. (45) to the quaternion form of the standard
Maxwell's equations [16,19].

Under the Lorentz transformations, the matrix $X$ is transformed
as
\begin{equation}
X^{\prime }=L^{+}XL ,\hspace{0.3in}L\in SL(2,c) , \label{47}
\end{equation}
where $L^{+}$ is Hermitian conjugated matrix. The matrices $\nabla
$, $\overline{\nabla}$, $F$, $A$ and $J$ are transformed as
follows
\[
\nabla ^{\prime }=L^{+}\nabla L ,\hspace{0.3in}F^{\prime
}=L^{-1}FL,\hspace{0.3in}J^{\prime }=L^{+}JL ,
\]
\vspace{-8mm}
\begin{equation}
\label{48}
\end{equation}
\vspace{-8mm}
\[
\overline{\nabla}^{\prime}=L^{-1}\overline{\nabla}\left(L^{+}\right)^{-1}
, \hspace{0.3in}A^{\prime }=L^{+}AL .
\]
The terms in Eq. (45), including the parameter ${\bf \theta}$
violate the Lorentz symmetry. The Lorentz-invariants of the
transformations (47), (48) are the determinants of matrices. The
spin-tensor formulation of the NC Maxwell's equations in the form
of Eq. (45) is convenient for considering the symmetric properties
of fields.

Let us consider some spin-tensor expressions in classical
electrodynamics when $\theta=0$. The energy-momentum tensor in the
case $\theta=0$ can be represented as follows
\begin{equation}
\tau _\beta T^{(\theta=0)}_{\beta \gamma }=\frac 12F^{+}\tau
_\gamma F , \label{49}
\end{equation}
where $F=(E_m +iB_m)\tau_m$. Using the field equation (45), it is
easy to prove that $\partial _\gamma T^{(\theta=0)}_{\beta \gamma
}=0$ at $J=0$. From Eq. (49) we verify that
\[
T^{(\theta=0)}_{44}=\frac 12\left( {\bf E}^2+{\bf B}^2\right)
,\hspace{0.3in} T^{(\theta=0)}_{k4}=i\epsilon _{kmn}B_mE_n ,
\]
\vspace{-8mm}
\begin{equation}
\label{50}
\end{equation}
\vspace{-8mm}
\[
T^{(\theta=0)}_{kn}=E_kE_n+B_kB_n-\frac12 \delta _{kn}\left( {\bf
E}^2+{\bf B}^2\right) ,
\]

so, the expression $T^{(\theta=0)}_{\beta \gamma }$ corresponds to
Eq. (34) at $\theta=0$. We may define the density of the Lorentz
force as
\begin{equation}
K_\alpha =\partial _\beta T^{(\theta=0)}_{\alpha \beta } .
\label{51}
\end{equation}
Then with the help of the matrix equation (45) at $\theta=0$ we
arrive at the relation
\begin{equation}
\tau _\alpha K_\alpha =-\frac 12\left( J^{+}F+F^{+}J\right) .
\label{52}
\end{equation}
Using Eq. (52) one can verify that the four-force $K_\alpha $
coincides with the known definitions:
\[
{\bf K}=\rho {\bf E}+{\bf J}\times {\bf B} ,
\]
\vspace{-8mm}
\begin{equation}
\label{53}
\end{equation}
\vspace{-8mm}
\[
K_4=i({\bf J}\cdot{\bf E}) .
\]
The expressions (49), (52) are form-covariant under the Lorentz
transformations due to Eq. (47).

\section{Conclusion}

We have just considered the Maxwell theory on NC spaces which is
described by the non-linear equations of the electromagnetic
fields. This means that the vacuum of electrodynamics on NC spaces
is similar to a medium with complicated (non-linear) properties.
The system of non-linear wave equations found, (22), (23),
possesses the solutions in the form of plane electromagnetic
waves. It was noted that the dual transformations of
electromagnetic fields under consideration are broken.

We have found the density of the energy and momentum of the
electromagnetic fields on NC spaces, and the canonical and
symmetric energy-momentum tensors. The canonical energy-momentum
tensor (34) is conservative, but the symmetric energy-momentum
tensor (38), found by varying the action on the metric tensor, is
non-conservative because the action of electromagnetic fields in
the case of NC spaces is not a Lorentz-scalar. It was shown that
the traces of the canonical and symmetric energy-momentum tensors
do not equal zero, i.e., there is a trace anomaly. The trace
anomaly is related with the violation of the conformal invariance,
and is a consequence of the breaking of the Lorentz invariance in
NC spaces. This anomaly is absent in the case of the plane
electromagnetic waves. The field equations are also obtained in
the compact spin-tensor (quaternion) form. The Lorentz
transformations of fields in the matrix form have been considered.
The spin-tensor formulation of the NC Maxwell's equations is
useful for different applications, especially for studying the
symmetric properties of fields.

There are various phenomenological effects of the
non-commutativity of coordinates (1) (see [2,3,12,20]). If
space-time is indeed non-commutative on short distances, it may
effect cosmology and early universe physics. It is important
because cosmology can verify the theories which are beyond the
standard model of particle physics. Probably, the consideration of
the field theories on NC spaces may solve the problem of dark
energy from trans-Plankian physics.

\begin{center}
{\bf APPENDIX: quaternion algebra}
\end{center}

Pauli's $2\times 2$-matrices $\tau _k$ ($k=1,$ $2,$ $3$) obey the
following relations
\[
\tau _m\tau _n=i\epsilon _{mnk}\tau _k+\delta _{mn},
\]
\begin{equation}
\tau_\mu \overline{\tau }_\nu +\tau_\nu \overline{\tau}_\mu
=-2\delta_{\mu \nu},
 \label{54}
\end{equation}
\[
\tau _\mu =\left( \tau_k,\tau_4\right) ,\hspace{0.3in}
\overline{\tau}_\mu =\left( -\tau_k,\tau_4\right) ,
\]
where $\tau_4=i\tau_0$, $\tau_0\equiv I_2$ is the unit $2\times
2$-matrix. The quaternion algebra can be realized with the help of
the Pauli matrices; setting $e_4=\tau_0$, $e_k=i\tau_k$ and using
the properties (54) we obtain the quaternion algebra which is
defined by four basis elements $e_\mu =(e_k,e_4)$ [19]) with the
multiplication properties:
\[
e_4^2=1,\hspace{0.3in}e_1^2=e_2^2=e_3^2=-1,\hspace{0.3in}e_1e_2=e_3,
\]
\begin{equation}
e_2e_1=-e_3,\hspace{0.3in}e_2e_3=e_1,\hspace{0.3in}e_3e_2=-e_1,
\label{55}
\end{equation}
\[
e_3e_1=e_2,\hspace{0.3in}e_1e_3=-e_2,\hspace{0.3in}e_4e_m=e_me_4=e_m,
\]
where $m=1,$ $2,$ $3$, and $e_4=1$ is the unit element.

The complex quaternion (or biquaternion) $q$ is
\begin{equation}
q=q_\mu e_\mu =q_me_m+q_4e_4,  \label{56}
\end{equation}
where the $q_\mu $ are complex numbers. Using the laws of
multiplication (55), we find that the product of two arbitrary
quaternions, $q$, $q^{\prime }$, is defined by:
\begin{equation}
qq^{\prime }=\left( q_4q_4^{\prime }-q_mq_m^{\prime }\right)
e_4+\left( q_4^{\prime }q_m+q_4q_m^{\prime }+\epsilon
_{mnk}q_nq_k^{\prime }\right) e_m. \label{57}
\end{equation}

It is convenient to represent the arbitrary quaternion as
$q=q_4+{\bf q}$ (so $q_4e_4\rightarrow q_4$, $q_me_m\rightarrow
{\bf q}$), where $q_4$ and ${\bf q}$ are the scalar and vector
parts of the quaternion, respectively. With the help of this
notation, Eq. (57) can be rewritten as
\begin{equation}
qq^{\prime }=q_4q_4^{\prime }-({\bf q}\cdot{\bf q}^{\prime
})+q_4^{\prime }{\bf q}+q_4 {\bf q}^{\prime }+ {\bf q}\times{\bf
q}^{\prime } . \label{58}
\end{equation}
Thus the scalar $({\bf q}\cdot{\bf q}^{\prime}) =q_m q_m^{\prime
}$, and vector ${\bf q}\times{\bf q}^{\prime }$ products are parts
of the quaternion multiplication. It is easy to verify the
combined law for three quaternions: $\left( q_1q_2\right)
q_3=q_1\left( q_2q_3\right)$.

The operation of quaternion conjugation (hyperconjugation) denotes
the transition to
\begin{equation}
\overline{q}=q_4e_4-q_me_m\equiv q_4-\mathbf{q}, \label{59}
\end{equation}
so that the equalities $
\overline{q_1+q_2}=\overline{q}_1+\overline{q}_2$, $\overline{
q_1q_2}=\overline{q}_2\overline{q}_1$ are valid for two arbitrary
quaternions $q_1$ and $q_2$. The modulus of the quaternion $\mid
q\mid $ is defined by $ \mid q\mid
=\sqrt{q\overline{q}}=\sqrt{q_\mu ^2}$. This formula allows us to
divide one quaternions by another, and thus the quaternion algebra
includes this division.

Quaternions are a generalization of the complex numbers and we can
consider quaternions as a doubling of the complex numbers. They
are convenient for investigating the symmetry of fields and
relativistic kinematics. In particular, the finite transformations
of the Lorentz eigengroup are given by [19]:
\begin{equation}
x^{\prime }=Lx\overline{L}^{*} ,  \label{60}
\end{equation}
where $x=x_4+{\bf x}$ is the quaternion of the coordinates
($x_4=it$, $t$ is the time, $x_m$ are the spatial coordinates),
$L$ is the quaternion of the Lorentz group with the constraint
$L\overline{L}=1$, $\overline{L}^{*}=L_4^{*}-{\bf L}^{*}$ and $*$
means the complex conjugation. The biquaternion $L$ with the
constraint $L\overline{L}=1$ is defined by six independent
parameters which characterize the Lorentz transformations. The
squared four-vector of coordinates, $x_\mu ^2$, is invariant under
the transformations (60): $x_\mu^{\prime 2}=x^{\prime
}\overline{x}^{\prime }=Lx\overline{L}^{*}L^{*}
\overline{x}\overline{L}=x\overline{x}=x_\mu ^2$, as
$\overline{L}^{*}L^{*}=L^{*}\overline{L}^{*}=1$. This shows that
the $6-$parameter transformations (60) belong to the Lorentz group
$SO(3,1)$.

\end{document}